\documentclass[nofootinbib,a4paper,11pt]{revtex4}
\usepackage{graphicx}

\usepackage{amssymb}
\usepackage{amsmath}
\usepackage{graphics}
\usepackage{color}
\usepackage[T1]{fontenc}
\usepackage{times,mathptm}
\usepackage[cp1250]{inputenc}
\usepackage{epsfig,epsf}
\newcommand{\be}{\begin{equation}}
\newcommand{\ee}{\end{equation}}

\newcommand\beq{\begin{eqnarray}}
\newcommand\eeq{\end{eqnarray}}
\begin{document}

\title{Remark on in-medium masses of hadrons with one heavy quark}

\author{M.~Sadzikowski}
\email{mariusz.sadzikowski@uj.edu.pl}
\affiliation{Institute of Physics, Jagiellonian University, Reymonta 4, 30-059
Krak\'{o}w, Poland
}

\begin{abstract}
In this short note we derive a formula which describes the
dependence of the mass of a hadron which contains a single heavy quark
on the temperature of the heat bath. It takes a simple scaling form
with the exponent which is different than in the case of the light hadrons.
The derivation is based on dimensional arguments within the framework of
the bag model paradigm. The simple realization of this scenario is presented
for the MIT bag model. The mass splitting between pseudoscalar and vector mesons
($D, D^\ast$ or $B, B^\ast$) as a function of temeperature is presented.
\end{abstract}

\maketitle

\section{Introduction}

It is a widely accepted conjecture that the properties of a hadron
change once it is placed in a hot and/or dense medium (for a recent review
see \cite{hayano}). It is a fundamental property of the physical vacuum, namely the spontaneous
symmetry breaking of the chiral symmetry and the deconfinement
transition, that lies at the foundation of this phenomenon \cite{brown}. The
changes of the vacuum influence hadron structure. All of these
aspects are manifestations of a non-perturbative nature of QCD,
therefore, challenge our understanding of the strongly interacting
matter.

Hadrons containing a single heavy quark create particularly simple
systems which allow to draw general conclusions and simple scaling relations.
Additionally, the properties of mesons with an open charm (as e.g. $D,D^\ast$) can influence
the evolution of the $J/\Psi$ particles in finite temperature strongly interacting medium
(see \cite{sibirtsev} or a recent paper \cite{blaschke}).

In this article, we try to describe the dependence of the mass of a hadron containing
a single heavy quark on temperature.
It occurs that basic assumptions and dimensional arguments are almost enough to
find a simple scaling formula. The essential simplification appears in the limits
of the massless quarks, when one considers light hadrons, and the
infinite mass quarks, when one is interested in the properties of heavy hadrons.
These limits supplemented with basic assumptions about the property of the vacuum
lead to the final formulae.

\section{Scaling}
It is a basic paradigm of the bag model \cite{mit_bag} that a hadron
is a bubble of the trivial, perturbative vacuum immersed in the complicated
physical vacuum. The true vacuum exerts
the pressure $B$ on the surface of the bag which is balanced by the quarks and gluons
confined inside the bubble. This conjecture is still an open question. It gives
a radius of the bag $R$ the physical meaning, even so we do not have at our disposal
a general definition of this quantity at the moment. This is exactly in opposite to the hypothesis
called "the cheshire cat principle" \cite{nielsen} which deprives the bag radius of the physical
meaning. This very neat assumption is well established in 1+1 dimensional space-time,
however, it is not clear whether it is still correct in 3+1 dimensions.
In this paper we assume that $R$ is the physical quantity which
provides us with the dimensional scale. The simplest phenomenological equation describing
the mass of the hadron has a form \cite{mit_bag2}
\be
\label{mass}
M=\frac{A}{R}+\frac{4}{3}\pi B R^3,
\ee
where the bag is spherically symmetric and $A$ is a parameter which depends on the
type of the hadron. This formula assumes implicitly that the considered quarks are massless.
This is an important assumption because the non-zero quark masses introduce the new scales
in the problem. Even within the MIT bag model phenomenology the above formula is not
correct if quarks are massive. The parameter $A$ contains many ingredients, in particular, the kinetic
energy of quarks and colour interactions. Let us also
notice that the formula (\ref{mass}) neglects the contribution from the surface term.
We think, however, that there is no point to attribute an independent physical meaning
to the surface degrees of freedom.
The radius of the bag is determined by the minimum energy principle $R=(A/4\pi B)^{1/4}$
which finally gives us the well known equation \cite{mit_bag2}
\be
\label{mass2}
M=\frac{4}{3}A^{3/4}(4\pi B)^{1/4} .
\ee
Let us notice that the parameter $A$ does not depend on temperature because of dimensional reasons. Indeed,
$A$ is a function of quantity $m R$, where $m$ is a light quark mass, the only dimensionless parameter in the model
which vanishes in the massless limit.

Now, let us turn our attention to the fact that the vacuum state of the QCD is
a function of the temperature and density \cite{polyakov}. This
phenomenon is the consequence of the asymptotic freedom and the non-abelian nature
of the colour fields. One could then think that the bag constant is in fact a temperature
dependent parameter\footnote{The bag constant depends also on the baryon chemical potential,
however, in this work we concentrate on the zero density line.}  which vanishes at the critical point $T_c$
\cite{rafelski}. This temperature is identified with the deconfinement transition. There has been a lot of work
done to find this dependence using simple
or more refine treatments. Let us mention, as an example, the approaches based on the Savvidy
ground state properties \cite{rafelski}, effects of coloured quark and ground state entropies \cite{miller},
counting of the number of the degrees of freedom \cite{pisarski}
or the chiral model calculations \cite{pisarski,song}. All the approaches could be summarized in
the equation $B(T)=B(1-f(T))$ where $f(T_c)=1$ and the exact form of the function $f(T)$
depends on the model.

Let us consider a single bag inside the vacuum at a non-zero temperature. Then at the thermodynamic
equilibrium the formula (\ref{mass2}) still holds but with the bag constant replaced with
the temperature dependent quantity $B(T)$. One can then discover the scaling relation \cite{pisarski}
\be
\label{scaling1}
\frac{M(T)}{M(0)}=(1-f(T))^{1/4}
\ee
with the critical exponent 1/4. This scaling has the same form for both baryons and mesons.

However, if one considers the hadrons containing a single heavy quark the situation becomes
different. The formula (\ref{mass}) does not hold anymore because we have an additional
energy scale at our disposal - the mass of the heavy quark. Using the expansion
in the inverse powers of the heavy quark mass one has \cite{shuryak,sad}
\be
\label{mass_H}
M_H=m_Q+\frac{A}{R}+\frac{4}{3}\pi B R^3+\frac{C}{m_QR^2} + O(1/m_Q^2).
\ee
The first term describes the
contribution coming from the heavy quark mass, the second and third ones are mass independent and
have the same form as the formula (\ref{mass}). The last term possesses simple
physical interpretation. It consists of two contributions one describing the residual
motion of the heavy quark and the other describing the interaction between the magnetic moment of the heavy quark
(which scales as $1/m_Q$) and the chromo-magnetic field of the light quarks. The magnetic contribution
depends on the spins of the quarks contained inside the hadron.
Assuming that $m_Q$ dominates the other scales, one can still keep
the relation $R=(A/4\pi B)^{1/4}$ neglecting $1/m_Q$ corrections.
It is very convenient now to consider the mass differences $\Delta M=M_{H}-M_{H^\prime}$
as a function of temperature. A particularly useful quantity is a difference  between pseudoscalar
and vector mesons. In that case the first three terms of (\ref{mass_H}) cancel and the last term
survives because it depends on the spin structure of the hadrons. Thus, we arrive at the scaling law
\be
\label{scaling2}
\frac{\Delta M(T)}{\Delta M(0)}=(1-f(T))^{1/2}.
\ee
One expects that
the corrections to the formula (\ref{scaling2}) behave as $O(1/m_Q^2)$ and are suppressed in the
limit of the infinite heavy quark mass. Let us notice that the exponent in equation (\ref{scaling2})
for the light hadrons is different and equal 1/4.

Such simple formulae (\ref{scaling1}, \ref{scaling2}) do
not hold for the strange quark which is neither massless nor very heavy.
In this case the masses difference is a function of these two scalings, however, its exact
shape is essentially unknown.

\section{MIT bag model realization}

We demonstrate the results derived in the previous section within the MIT bag model\footnote{It is worth to mention
that the MIT bag model was successfully applied in the relation between the hadron gas model and lattice QCD thermodynamics \cite{karsch}.}.
The mass formula (\ref{mass_H}) for mesons can be written as \cite{shuryak,sad}
\be
\label{mass_HQ}
M=m_Q+\frac{x-Z}{R}-\frac{4\alpha_s}{3R}\left(C+\frac{1}{2}\right)+\frac{4}{3}\pi B R^3
+\frac{x^2}{2m_QR^2}+ E_M
\ee
where $x/R$ is a momentum of the light quark and according to the MIT bag phenomenology $x=2.04$ for the massless
quark. The parameter $Z$ describes the Casimir energy of the closed cavity. The third term follows from
the chromo-electric interactions between quarks. The residual motion of the heavy quark is responsible for the fifth term
which has a form of the non-relativistic kinetic.
The chromo-magnetic interaction is equal to \cite{shuryak,sad}
\be
\label{Em}
E_M=\frac{D\alpha_s}{m_QR^2}\langle\vec{\sigma}_Q\vec{\sigma}_q\rangle
\ee
where the spin-spin interaction term
$\langle\vec{\sigma}_Q\vec{\sigma}_q\rangle = \{-3,1\}$ for pseudoscalar and vector mesons respectively\footnote{The parameters $C=0.52$ and $D=1.04$ for the massless light quarks \cite{sad}.}.
A delicate point is connected with the
strong coupling constant $\alpha_s$. In principle it could be also the function
of the bag radius but then it would only give us a weak logarithmic corrections to the scaling
(\ref{scaling2}). Thus the mass difference between the pseudoscalar and vector mesons has a form
\be
\Delta M=\frac{4 D\alpha_s}{m_QR^2}
\ee
which leads exactly to the scaling given by equation (\ref{scaling2}) with the exponent $w=1/2$
once $R\sim B^{-1/4}$.

The comparison between scalings for the light and heavy hadrons is shown in Fig. 1. As an example, one
can consider the function $f(T)$ calculated in the paper \cite{song}
\be
\label{Tdependence}
f(T)=B_0\left(1-\frac{\delta}{B_0^{1/2}}\left(\frac{T}{B_0^{1/4}}\right)^2+\beta\left(\frac{T}{B_0^{1/4}}\right)^4\right)
\ee
where $\delta = 46.039$ MeV$^2$ and $\beta = 3.016$. For the purpose of the presentation
it is enough to take $(B_0)^{1/4} = 224$ MeV which gives the critical temperature $T_c=170$ MeV\footnote{The temperature scale
is set by the number of light quarks $N_f=2,3$ because the heavy quarks rather play a role of external
color sources. In such situation the crossover temperature from lattice simulations is close to 170 MeV within the errors e.g. \cite{lattice}.}. One can
also notice that from equations (\ref{mass_HQ},\ref{Em}) follows that the masses of pseudoscalar meson
increases whereas for vector meson decreases with increasing temperature.

\begin{figure}
\centerline{\epsfxsize=10 cm \epsfbox{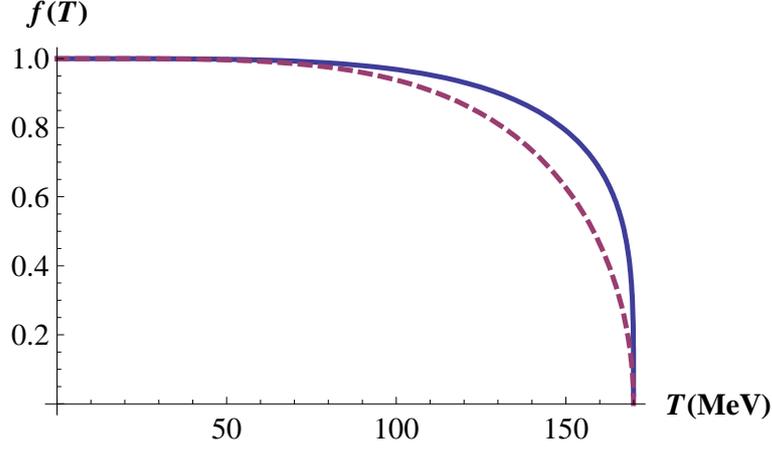}}
\caption{Comparison of the scaling formulae: (\ref{scaling1}) - the thick line, and (\ref{scaling2}) - the dashed line
as a function of temperature given by equation (\ref{Tdependence}).}
\end{figure}


\section{Conclusions}

We have shown that the dependence of the hadron masses on temperature depends on
the flavour content of the hadron. There are two different scaling regimes one
in the limit of the massless quarks and another one in the limit of the infinite quark mass
$m_Q\rightarrow \infty$.
If one keeps the corrections of the order of $1/m_Q$ then the formula (\ref{scaling2}) holds.
From the other side, the massless limit leads to the formula
(\ref{scaling1}). For the non-zero masses, as in the case of the strange quark,
non of the above formulae are correct. We do not find any model independent way
to derive a simple form of the scaling in this situation.

In our opinion the scaling form (\ref{scaling2}) is rather of
theoretical than phenomenological interest and can be used as a test
of our understanding of the physics of the light part of the heavy hadron.
It would be then interesting to check the discussed results again the model calculations.
One can consider the chiral models within
the scheme of the heavy quark effective theory \cite{neubert}. It is also possible
to try more refine bag models including the chiral or the soliton
bag model. Finally one could also think about the lattice calculations, however, the
heavy quark physics is still a difficult challenge for the computer simulations.

It is also worth to note that equations (\ref{mass_HQ},\ref{Em}) together with
relation $R=(A/4\pi B)^{1/4}$, where $B$ is given by formula (\ref{Tdependence}), can be used for the
direct estimation of the masses of mesons as a function of temperature. This is an interesting issue because
the masses of $D, D^\ast$ mesons directly influence
$J/\Psi$ evolution in hot (and dense) medium which is of prime interest for
the understanding of heavy-ion physics at LHC energies \cite{sibirtsev,blaschke}.
Similarly one can also expect similar influences between the behavior of
$B, B^\ast$ mesons and $\Upsilon$.

\end{document}